\documentclass[letterpaper]{sfchem}
\usepackage{graphicx}

\begin{document}

\title{Results from a VLT-ISAAC survey of ices and gas around young stellar objects}

\author{K. M. Pontoppidan\inst{1} \and H. Fraser\inst{1} \and F. L. Sch{\"o}ier\inst{1}  \and E. Dartois\inst{2} \and W.-F. Thi\inst{1,3}  \and E. F. van Dishoeck\inst{1}} 
  \institute{Leiden Observatory,  
  P.O.\ Box 9513, 2300 RA Leiden, The Netherlands \and Insitut d'Astrophysique Spatiale, B{\^a}t. 121, Universit{\'e} Paris XI, 91405 Orsay Cedex, France\and {Astronomical Institute "Anton Pannekoek", University of Amsterdam, Kruislaan 403, 1098 SJ Amsterdam, The Netherlands}} 
\authorrunning{Pontoppidan et al}
\titlerunning{A VLT-ISAAC survey of ices and gas}

\maketitle 

\begin{abstract}
General results from a $\rm 3-5~\mu m$ spectroscopic survey of nearby low-mass young stellar objects are presented. 
$L$ and $M$-band spectra have been obtained of $\sim50$ low mass embedded young stars using the
ISAAC spectrometer mounted on UT1-Antu at Paranal Observatory. For the first time, a consistent census of the CO, $\rm H_2O$ ices and the minor ice species $\rm CH_3OH$ and
$\rm OCN^-$ and warm CO gas present around young stars is obtained, using large number statistics and
resolving powers of up to $R=10\,000$. The molecular structure of circumstellar
CO ices, the depletion of gaseous CO onto grains in protoplanetary disks, the presence of hot gas in the inner parts of circumstellar disks
and in outflows and infalls are studied. Furthermore, the importance of scattering effects for the interpretation of the spectra have been addressed.

\end{abstract}

\section{Introduction}
The $\rm 3-5~\mu m$ region of young stellar objects contains a large range of spectroscopic signatures of both cold ices as well as
warm and hot gas. The most important are the broad absorption band at $\rm 3.08~\mu m$ due to water ice, the narrow absorption band
around $\rm 4.67~\mu m$ due to solid CO and the large number of lines from the fundamental ro-vibrational transitions from gaseous CO
covering the entire $M$-band. The ice mantles are thought to be located on dust grains in the cold midplanes of optically thick circumstellar disks, in the cold and dense envelopes of collapsing protostars and to some extent in the more quiescent parts of dense molecular clouds. However,
due to the nature of absorption line studies, it is often hard to disentangle contributions from different environments
along the line of sight. Studying large samples statistically is one way of circumventing the problem. Alternatively absorption studies can be combined with observations of rotational transitions and submm continuum to map the foreground clouds (\cite{ab00}).

The fundamental ro-vibrational transitions of CO \linebreak present in the $\rm 4.5-5.0~\mu m$ region probe gas of temperatures from 20 K to more than 
5000 K due to the large range of rotational and vibrational levels present in a narrow wavelength region and are thus sensitive temperature indicators.  Ro-vibrational transitions of various molecules in the mid-infrared region are among the few effective probes of warm and hot gas
located behind dense, dusty envelopes and are therefore a powerful probe of warm gas in the inner few AUs of circumstellar disks
and the hot gas in the region between the disk, the central star and the outflow.  

\section{Observations}
The observations were obtained between January 2001 and May 2002 using the Infrared Spectrometer and Array Camera (ISAAC) mounted
on UT1-Antu at Paranal Observatory. $L$-band spectra were taken at low resolving powers of $\lambda/\Delta\lambda=600-1200$ and
$M$-band spectra were obtained using the medium resolution mode, resulting in $\lambda/\Delta\lambda=5\,000-10\,000$.
An advantage of the ISAAC instrument is the ability to
rotate the detector to include two sources in the slit. This, combined with the often excellent seeing in the $\rm 3-5~\mu m$ region (typically 0\farcs3 to 0\farcs5), enabled us to obtain simultaneous spectra of several close binaries with separations of $1-5\arcsec$ to probe
the difference in ice chemistry and structure along lines of sight separated by only a few hundred AU. 
A total of $\sim50$ sources has been observed. The survey was designed to focus on low-mass 
($L=0.1-50~L_{\odot}$) class I and borderline class I-II 
sources in the major southern low-mass star-forming clouds, $\rho$ Ophiuchus, Corona Australis, Chameleon and Serpens. Some intermediate
to high-mass sources as well as some background stars were observed for comparison. Spectra of high quality of sources as faint
as $L=10$ and $M=8$ were obtained, thus probing the entire stellar mass range. Here we present recent results from the analysis
of the $M$-band ($\rm 4.5-5.0~\mu m$) spectra. An initial account is given by \cite*{ewine}, which includes the full list of team members.

\section{Summary of results}
\subsection{The solid CO absorption band profile}

\begin{figure*}[ht]
\centering
\includegraphics[width=0.8\linewidth]{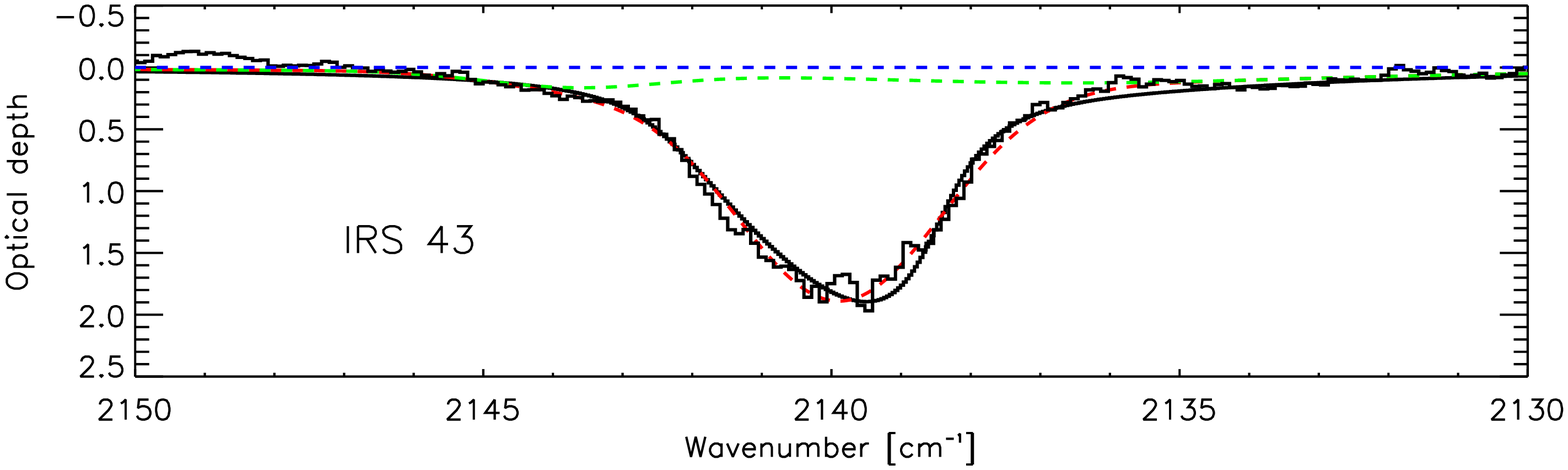}
\caption{Example of a fit to the solid CO profile using a simple physical model (Pontoppidan et al. 2003). The red curve shows
the total analytical model, the black solid curve shows the total physical model while the green curve shows the sum of the
blue and red analytical components.}
\label{irs43}
\end{figure*}

The profiles of the solid CO stretching vibration mode around $\rm 4.67~\mu m$ are analysed in \cite*{po03}. It is found that
the structure of the profile is almost identical along all observed lines of sight. A phenomenological decomposition is used to show that
the observed profiles can be split into three unique components. They have the same centers and widths, such that only their relative intensities need to be varied to reproduce all the observed circumstellar and interstellar CO bands. The three components are centered
at $\rm 2143.7~cm^{-1}$, $\rm 2139.9~cm^{-1}$ and $\rm 2136.5~cm^{-1}$. Using a physical model
of the CO ice, it is found that the two blue components can be fully explained by pure crystalline CO ($\alpha$-CO). The model
depends only on two free parameters, namely the number of CO oscillators per unit volume and the linear polarisation fraction of the background source. 

The identification of the third red component is less secure. It may be associated with CO molecules embedded on the water ice known to exist in large quantities in all
lines of sight also containing solid CO, although there are serious inconsistencies 
with the available laboratory spectra of water-rich ice mixtures. For instance, the $\rm 2152~cm^{-1}$ feature due to CO-OH bonds is not detected in any
of the observed spectra, but strong upper limits can be derived. This feature, in addition to a red $\rm 2137-2135~cm^{-1}$ feature, is always present in amorphous and unprocessed water-rich mixtures.  Substantial thermal or energetic processing will remove the $\rm 2152~cm^{-1}$  feature but will also irreversibly change the
shape of the $\rm 2136.5~cm^{-1}$ feature so that it no longer matches the astronomical spectra.  The optical depth of a broad absorption band centered on $\rm 2175~cm^{-1}$ is found to correlate strongly with the depth of the $\rm 2136.5~cm^{-1}$ band for low-mass stars only.
This feature may be an indication that another carrier is present in this region in addition to the $\rm 2165~cm^{-1}$ XCN band.  

An example of a fit using the
physical model of the CO ice is shown in Fig. \ref{irs43}. 
The similarity between the CO ices observed along a large variety of lines
of sight shows that the CO profile is not a good probe of chemical processing of circumstellar ices. The presence of secondary species
in the ice also cannot be inferred from the profile of the CO stretching vibration, contrary to what has been previously reported in the 
literature. The abundance of species such as $\rm N_2$, $\rm O_2$ and $\rm CO_2$ mixed with the CO ice is probably less than 10\%.
However, the presence of solid CO profiles which show only the red component suggests that the carrier of the red component is the least volatile of  three carriers. The ratio of the middle and blue components to the red component may then be a good indicator thermal processing. 
This supports the idea that the CO responsible for the red component is trapped in the water ice or in another environment. 

The presence of pure crystalline CO ice and the apparent low concentration of contaminants mixed with the bulk of the CO ice sets strong
constraints on the condition under which CO freezes out onto dust grains. 

\subsection{Gaseous CO ro-vibrational emission lines}
Luminous sources often exhibit deep absorption lines
sometimes showing signatures of infall or outflow in the gas. Low luminosity sources often show pure emission lines or self-absorbed emission lines. 
One extreme example which has been studied as an isolated case is GSS 30 IRS1 in the $\rho$ Ophiuchi cloud core (\cite{po02}). This source
shows narrow ($\rm <30~km~s^{-1}$) and bright emission lines from CO gas. The emission lines are well-fitted by gas at a single
temperature of $\rm 515\pm10~K$.  The gas is not seen directly and the lines are probably scattered into the line of sight on the walls of a bipolar cavity.  The fact that a range of temperatures does not seem to characterise the gas is surprising. The heating mechanism may be related to 
an accretion shock as circumstellar gas is incorporated into a circumstellar disk. A single temperature of $\rm \sim 500~K$ then arises naturally as 
the recombination of $\rm H_2$ behind the shock front releases chemical energy which maintains a thermal equilibrium (\cite{nh}).

\begin{figure}[ht]
\resizebox{\hsize}{!}{
\includegraphics[width=0.8\linewidth]{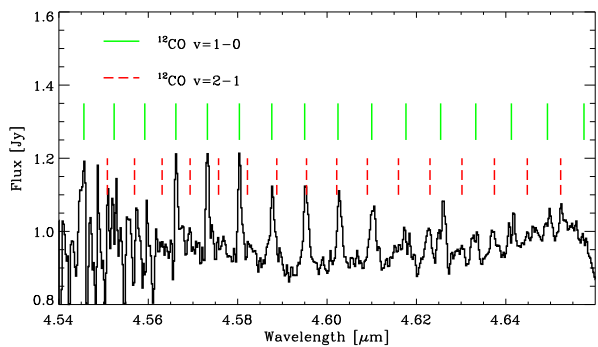}}
\caption{Example of broad gas phase CO lines in the class I source Elias 32 located in the $\rho$ Ophiuchi cloud (Pontoppidan et al. 2003).}
\label{irs43gas}
\end{figure}

 Other low-mass sources show very broad ($\rm >100~km~s^{-1}$) CO emission lines from hot ($T_{\rm gas}>1\,000~\rm K$) CO gas. An example of this is shown in Fig. \ref{irs43gas} for Elias 32. The origin of the different classes of ro-vibrational lines (broad versus narrow, emission versus absorption)
is presently unclear, although geometry and the presence or absence of foreground clouds likely plays a role. In general, studies employing full radiative transfer models are needed to constrain the detailed origin of the lines.

\section{Conclusions and future Work}
The combination of spectra of solid state features taken at the highest possible spectral resolution combined with physical models
of the molecular structure and careful laboratory experimentation, also using high resolution spectrometers, is a powerful tool for the  
study of the microscopic structure and composition of interstellar solid state material. Furthermore, high resolution spectroscopy of the ro-vibrational 
transitions of molecules is well suited for penetrating the obscuring envelopes of young embedded stars to study the physics and 
relationships of the inner parts of circumstellar disks, accretion of circumstellar material and the inner engines of outflows. Detection of
strong solid CO absorption through an edge-on disk has recently been published by \cite*{wf}.

Work in progress includes:
\begin{itemize}
\item Full radiative transfer modeling of the observed ro-vibrational gas-phase CO lines to constrain the origin of the different classes (narrow absorption
lines, narrow emission lines and broad emission lines).
\item Using the statistics of the large sample to analyse the relationship between solid CO and water ice in interstellar grain mantles.
\item Performing laboratory experiments to determine the origin of the $\rm 2175~cm^{-1}$ feature and its relation to the red component
of the CO ice profiles.
\item Further modeling of the formation of ice mantles to explain the observed icy features. Emphasis will be put on
determining the detailed structure of the mantles, including the degree to which the ices are expected to be mixed, 
the composition as a function of the depth in a single grain mantle and the evolution of the mantle during the formation
of a low-mass star.
\end{itemize}

\begin{acknowledgements}

The authors wish the thank the VLT staff for assistance and advice, in particular Chris Lidman for many helpful comments
on the data reduction. This research was supported by the Netherlands Organization for Scientific Research (NWO) grant 614.041.004, the
Netherlands Research School for Astronomy (NOVA) and a NWO Spinoza grant.
\end{acknowledgements}

\end{document}